Research Article

# Modeling Competitive Dynamics with Heterogeneous Lanchester Equations: Insights from Kursk Battle

Sumanta Kumar Das*

IASRI, IARI, New Delhi, India

Email: sumantadas.delhi@gmail.com (Sumanta Kumar Das)

**ORCID**

https://orcid.org/ 0000-0003-1501-7000 (Sumanta Kumar Das)

**Abstract**

This study extends the classical generalized homogeneous Lanchester model for heterogeneous situation. The complexity of parameter estimation in heterogeneous situation is handled through statistical parameter estimation method along with simulation tool. Goodness of fit measures of these parameters are further improved by strategically dividing the entire operation in multiple stages. This study will enhance both the parameter estimation of nonlinear combat process as well as system dynamics involving multiple factors. The relative comparisons of attrition coefficients and exponent parameters revel the hidden information in the data which may not be possible to visualize in the homogeneous situation. The study is also useful for analytical based wargame development which may not be possible with detailed modelling and simulation approaches.

## 1. Introduction

Lanchester [10] based models of combat process have gained significant importance in the 20th century. These equations are widely used for modeling in warfare and representing the decrease in force levels over time commonly referred to as attrition. Lanchester in 1914 proposed a set of differential equations, which quantify the importance of force concentration on the battlefield. Many authors have subsequently modified his original work to represent combat dynamics in modern warfare. The features of the Lanchester equation that makes it suitable for analysis includes:

*Applicability in the competitive world*: Lanchester models are widely used for historical battle analysis. Other than analysing human warfare Lanchester model have also been used for analysis of fights among social animals, market analysis [30].

*Force Aggregation*: Lanchester models are found to be suitable for developing aggregated combat modeling using High Resolution Simulation Model. In reality, actual historical combat data is not easily available and common practice is to develop High Resolution simulation data with detailed design. Various literatures [24, 25] have demonstrated that estimating attrition rates from high-resolution simulation and using Lanchester model for linking the various resolution of different simulation model.

*Flexibility with different resolutions*: Lanchester models are flexible for both homogeneous as well as heterogeneous situations. Lanchester models are used for theoretically consistent force aggregation and disaggregation in two dimensions [16].

Regardless of credits of prior discovery, Lanchester's equations are used worldwide for calculating attrition rates. We propose a general form of the heterogeneous Lanchester's model as:

$$\dot{X}_{i'_t} = \sum_{i=1}^{F} a_i \, {X_{i'_t}}^{q_i} Y_{i_t}^{p_i} \qquad (1)$$

$$\dot{Y}_{i'_t} = \sum_{i=1}^{F} b_i Y_{i'_t}^{q_i} X_{i_t}^{p_i}, \quad \forall i' = 1,2,\ldots,F \quad \textbf{(2)}$$

Where $X_{i_t}$ denotes the strengths of the i[th] type of Red forces at time *t* and $Y_{i_t}$ denotes the strength of the $i^{th}$ type of Blue forces at time *t*. $\dot{X}_{i'_t}$ and $\dot{Y}_{i'_t}$ are red and blue forces killed at time *t*.

$a_i$ represents attrition rate of *i*th type of Blue forces and

$b_i$ represents attrition rate of *i*th type of Red forces;

$i$=1,2,…$F$ where $F$ denotes the total number of forces.

$p_i$ is the exponent parameter of the attacking force,

$q_i$ is the exponent parameter of the defending force.

Equations (1) and (2) involve unknown parameters ai, bi, pi and qi. We are generally acquainted with two forms of these equations for homogeneous weapon engagement (when $i$=1). Lanchester linear law in which$p_i = q_i = 1$and force ratios remain equal if $a_i X^{p_i}{}_{i_0} = b_i Y^{p_i}{}_{i_0}$. Lanchester's linear law is interpreted as a model from a series of one-on-one duel between homogeneous forces and this law describes combat under "ancient conditions". The equation is also considered a good model for area fire weapons, such as artillery. Lanchester square law in which $p_i = 0,\ q_i = 1$ , that is, force ratios remain equal if $(a_i X^{p_i}_{i_0})^2 = (b_i Y^{p_i}_{i_0})^2$ applied to modern warfare in which both sides are able to aim their fire or concentrate forces.

On integrating equation (1) and (2) we obtain the state equation:

$$\frac{Y^{p_i}_{i_0} X^{q_i}_{i'_0} - Y^{p_i}_{i_t} X^{q_i}_{i'_t}}{X^{p_i}_{i_0} Y^{q_i}_{i'_0} - X^{q_i}_{i_t} Y^{p_i}_{i'_t}} = \frac{b_i}{a_i} \qquad (3)$$

here $X_{i_0}$, $Y_{i_0}$ represent the initial values of Blue and Red forces respectively. This equation says that the relationship between the power of the losses in any fixed time period is equal to the inverse ratio of the attrition rate parameters. Equation (3) leads to the victory condition for Blue. Most forces have breakpoints at which they will cease fighting and either withdraw or surrender if:

$$\frac{Y^{p_i}_{i_0} X^{q_i}_{i'_0} - Y^{p_i}_{i_t} X^{q_i}_{i'_t}}{X^{p_i}_{i_0} Y^{q_i}_{i'_0} - X^{q_i}_{i_t} Y^{p_i}_{i'_t}} \geq \frac{b_i}{a_i} \qquad (4)$$

Finally equations (4) may be solved in closed form as function of t.

$$X^{p_i}_{i'_t} = \frac{1}{2}\left(\left[X^{p_i}_{i_0} - Y^{q_i}_{i_0}\sqrt{\frac{a_i}{b_i}}\right] e^{t\sqrt{a_i b_i}}\right)$$

$$+\frac{1}{2}\left(\left[X^{p_i}_{i_0} + Y^{q_i}_{i_0}\sqrt{\frac{a_i}{b_i}}\right] e^{-t\sqrt{a_i b_i}}\right)$$

$$Y^{q_i}_{i'_t} = \frac{1}{2}\left(\left[Y^{q_i}_{i_0} - X^{p_i}_{i_0}\sqrt{\frac{b_i}{a_i}}\right] e^{t\sqrt{a_i b_i}}\right)$$

$$+\frac{1}{2}\left(\left[Y^{q_i}_{i_0} + X^{p_i}_{i_0}\sqrt{\frac{b_i}{a_i}}\right] e^{t\sqrt{a_i b_i}}\right) \qquad (5)$$

There is another form of mixed combat model where attacker uses area fire ($p_i = 1, q_i = 1$i.e. linear form) against a defender using aimed fire ($p_i = 1, q_i = 0$, i.e. square form). This mixed form of combat model is known as ambush model proposed by Deitchman [5].

Early empirical applications of **Lanchester models** focused primarily on testing whether the classical homogeneous formulations could explain historical combat outcomes. Helmbold's analysis [26] of the Iwo Jima campaign and Bracken's study [1] of the Ardennes offensive demonstrated that one-sided or homogeneous Lanchester equations could be fitted to real battle data, though often with simplifying assumptions. Subsequent work by Clemens [2], and later Lucas and Turkes [11], extended these efforts to the **Kursk campaign**, again relying largely on homogeneous formulations and least-squares estimation. Willard [20] has tested the capability of the Lanchester model for analyzing the historical battle data for the battles fought between the years of 1618-1905. NR Johnson and Mackey [31] analyzed the Battle of Britain using the Lanchester model. This was a battle of an air combat between German and Britain. Wiper, Pettit and Young [21] applied Bayesian computational techniques to fit the Ardennes Campaign data.

Several studies revisited Bracken's Ardennes results [1]. Fricker [7] introduced logarithmic transformations and regression-based extensions, concluding that neither the linear nor square laws adequately captured the dynamics of the campaign. Hartley and Helmbold [8], analyzing the Inchon–Seoul campaign, similarly found that constant-coefficient square-law models performed poorly unless the battle was subdivided into smaller engagements. Their findings underscored a recurring limitation: classical Lanchester laws often fail to generalize across battles or even across phases of the same battle.

Parallel developments explored alternative estimation frameworks. Wiper, Pettit, and Young [21] applied Bayesian computational methods to the Ardennes data, showing that stochastic formulations and prior information can yield parameter estimates that differ substantially from deterministic least-squares results. Their work highlighted the sensitivity of Lanchester fits to both model structure and estimation technique.

Research on the Kursk campaign continued with Turkes [18], who emphasized the scarcity and asymmetry of historical data and demonstrated that a wide variety of Lanchester variants could fit the Kursk dataset. Lucas and Turkes [11] later introduced a parameter-search approach to evaluate how different exponent and attrition-rate combinations explain the Kursk and Ardennes battles. Their conclusion—that none of the classical linear, square, or logarithmic laws consistently fit the data—reinforced the need for more flexible formulations.

More recent studies have broadened the scope of Lanchester-type modeling. Cangiotti et al. [35] generalized unaimed-fire models to multi-force engagements, while Fifelola et al. [36] used Jacobian linearization and energy-based stability analysis to examine asymmetric warfare dynamics. Ji et al. [37] integrated Lanchester laws with Nash equilibrium concepts to model autonomous swarm confrontations. Collectively, these contemporary contributions point toward the importance of **heterogeneous modeling** and multi-factor attrition processes in modern and historical contexts.

The classical attrition theory is further extended in the study [41], it introduces spatially varying formulations that allow localized solutions, how combat outcomes vary across different regions of engagement. This work highlights the importance of incorporating spatial heterogeneity into Lanchester-type models, moving beyond aggregate force interactions to more nuanced, location-dependent dynamics.

Across these studies, three themes emerge. First, most historical analyses rely on homogeneous Lanchester equations, even when battles involve multiple interacting weapon systems. Second, parameter estimation has been dominated by **least-squares methods**, with limited exploration of alternative statistical frameworks such as **maximum likelihood estimation**. Third, goodness-of-fit assessments often depend on a single metric—typically the sum of squared residuals—providing an incomplete picture of model adequacy.

These gaps motivate the present study, which fits both homogeneous and heterogeneous Lanchester formulations to the Kursk campaign using two estimation approaches (LSE and MLE) and evaluates model performance using a broader suite of goodness-of-fit measures, including Kolmogorov–Smirnov, chi-square, and $R^2$. By incorporating artillery effects and comparing estimation landscapes through contour and 3D surface visualizations, this work contributes a more comprehensive assessment of how heterogeneous Lanchester models perform when applied to detailed historical battle data.

Recent advances in data-driven modelling highlight the importance of integrating multiple risk factors into predictive frameworks. For example, El-Gezawi (2025) [39] developed a data-driven system for intelligent traffic risk prediction and accident prevention, combining statistical estimation with real-world data to improve accuracy. This approach parallels the methodological challenges in combat modelling, where heterogeneous forces (tanks and artillery) jointly influence attrition outcomes. Just as traffic risk prediction requires accounting for multiple interacting variables, heterogeneous Lanchester models must incorporate cross-effects between weapon systems. The emphasis on maximum likelihood estimation and goodness-of-fit validation in El-Gezawi's work provides a useful analogue for our own comparison of estimation techniques in modelling the Kursk battle.

Recent work in applied operations research has emphasized the value of incorporating soft computing and fuzzy set theory to model complex, uncertain systems. For instance, Jahanvi and Singh (2026) [40] explored smart clothing innovations in human augmentation using fuzzy set theory to capture imprecision and variability in human–technology interactions. Their approach demonstrates how non-linear, multi-factor systems can be represented more realistically when uncertainty is explicitly modelled. This insight is directly relevant to heterogeneous Lanchester modelling, where the interaction between tanks and artillery introduces uncertainty in attrition rates. By analogy, fuzzy and probabilistic frameworks can enhance the robustness of combat models, allowing them to better reflect the variability of battlefield conditions.

The aim is to present a **framework for competitive dynamics under uncertainty** using heterogeneous Lanchester equations. Key parameters, namely **attrition coefficients** and **exponent values**, capture the effectiveness of heterogeneous resources such as advertising, pricing, and innovation. We employ two standard methods: **Least Squares Estimation (LSE)**, and **Maximum Likelihood Estimation (MLE**. LSE offers computational simplicity, while MLE provides statistically efficient estimates. Together, they form a robust basis for analyzing **multi-force competition** and guiding resource allocation under uncertain conditions.

In the next section we have discussed in detail the mathematical formulations of homogeneous and heterogeneous situations. We have seen in Bracken [1], Fricker [7], Clemens [2], Turkes [18], Lucas and Turkes [11] that LSE method have been applied for evaluating the parameters for fitting the homogeneous Lanchester equations to the historical battle data. The MLE method [15-16] has not been explored particularly for fitting the historical battle data till date. Also, only one measure i.e. Sum-of-squared-residuals (SSR) has been explored for measuring the Goodness-of-fit (GOF). The main objective of this study is to assess the performance of the MLE approach for fitting homogeneous as well as heterogeneous Lanchester equations to the Battle of Kursk. Various measures of GOF [4] viz. Kolmogorov-Smirnov, Chi-square and $R^2$ have been computed for comparing the fits and to test how well the model fits the observed data. Applying the various GOF measures considering the artillery strength and casualties of Soviet and German sides from the Kursk battle data of World War-II validates the performance of MLE technique.

Section 2 presents in brief the overview of the battle of Kursk. Section 3 describes the mathematical formulation of likelihood estimation in case of both homogeneous as well as heterogeneous situations. Section 4 describes the Tank and Artillery data of Battle of Kursk and discusses the methodology for implementing the proposed as well as other approaches. Also, this section contains a performance appraisal of the MLE using various GOF measures. Section 5 analyses the results after observing various tables and figures and discusses how well the MLE fits the data. Section 6 summarizes the important aspects of the paper.

## 2. HISTORY OF THE BATTLE OF KURSK

After suffering a terrible defeat at Stalingrad in the winter of 1943, the Germans desperately wanted to regain the initiative. In the spring of 1943, the Eastern front was conquered by a salient, 200 km wide and 150 km deep, centred on the city of Kursk. The Germans planned in a classic pincer operation named Operation Citadel, to eliminate the salient and destroy the Soviet forces in it. On 2 July 1943, Hitler declared, "This attack is of decisive importance and it must succeed, and it must do so rapidly and convincingly. It must secure for us the initiative.... The victory of Kursk must be a blazing torch to the world."[31]

The Germans started the Battle of Kursk on July 4, 1943 on the southern half of the Kursk salient, but this was merely to gain better artillery observation points. The battle began in earnest early in the morning of July 5, when Soviets conducted an artillery barrage before the

Wehrmacht attacked. The Germans countered with their own planned barrages shortly thereafter and seized the initiative on both fronts. Soviet General Rokossovsky redeployed his reserves on the night of July 5 in order to attack the following day. The divisions of the 17th and 18th Guards

Rifle Corps, with support from the 3rd, 9th, 16th, and 19th Tank Corps, were beginning offensive operations at 5:30 a.m. on July 6 in support of 13th Army. July 6 was considered as worst single day of CITADEL for German tank losses. On 7th July the German attack with armour forces in the northern and northeast side and captured the village Lutschki and continued advancing towards the village of Tetrevino against the very strong Soviet infantry and armor battle. By evening July 7, the Germans were ableto capture the village Tetrevino. On 8th July the German attacked with armour forces and barely captured the Teploye village. The Soviet counterattacked and recapture the lost Teploye village.The soviet defended very well on that day although they lost 315 tanks on that day in comparison to 108 tank loss of the Germans. The German forces wanted to develop a sharp wedge towards Kursk via Oboyan village. The German forces attacked with more than 500 tanks. The soviet forces defended the Oboyan with sophisticated artillery guns. Despite the strong defense the Germans were able to foothold over the Pena River. During the period of July 7-9both the sides had suffered largest number of tank losses. Due to this reason the Germans planned to attack from less resistance Prokorovka side. After changing the direction of the attack the German reached and seized the village of Novoselovka. The Soviets understood the German’s plan and they started using their reserve units. But despite of that the Germans were able to break through the Soviet defenses by evening of the July 10. The German intention at this point was to cross the river Psel to the extent of as many as troops and vehicles are possible. The Germans were able to seize a bridge. The engagement was at this point between artillery and tank. Both the forces were preparing for the battle of Prokhorovka. The Battle of Prokhorovka was the decisive phase of the Battle of Kursk. The Soviets started with artillery defense and later it turns out to be totally tank against tank meeting engagement. The German tanks had to face minefield as wellas well defended Soviet anti-tank weapons. The resulting titanic battle was a tactical draw. The Germans lost 98 tanks against 414 Soviet tank losses. Hitler called off the battle. The losses from the fighting over July 12 and 13 were extensive on both sides. In KUTUZOV there was a heavy armour engagement between the two forces. The German forces destroyed 117 Soviet tanks. The Soviets also damaged 57 German tanks. The battle on Prokorovka still continued on 14th July. The German planned an offensive operation named operation Roland. It started on July 14. The aim was to destroy the Soviet armor reservoirs. The German armor units fought with the artillery forces that were defending the armor reservoirs of the Soviet in the southern part of the Prokorovka. Several tactical positions were captured by the German forces. On this day the Germans were capable of performing minor offensive operations and they were launching attacks to form the Gostishchevo-Liski pocket. Hitler redirected it back to Isyum on July 16. According to the various war analysts it is being considered that because of Hitler's decision, von Manstein lost the availability of a powerful mobile formation that could have been very useful in the battle. During this time most of the engagements were between German infantry and Soviet tanks. Most of the damage was suffered by the Soviet because they were not equipped with modern antitank weapons that can deal effectively with the Soviet armour. The Soviets launched their counteroffensive along the Mius River on July 17. The Southwestern Front, commanded by Colonel General Tolbukhin, attacked the heavily fortified Mius River line defenses. The Soviet counter attack was known as operation RUMANTSYEV.

## 3. MATHEMATICAL FORMULATION

Let S denote the time between two consecutive casualties for a side, its probability density function is denoted by $f_S(s)$. Let $(m^i{}_k, n^i{}_k)$ represents the ith type force strengths (e.g. tank, artillery etc.) of blue and red forces of a battle for the kth time instance respectively. Let us also denote (for k=1,2…K) the time (a random variable) at which Kth casualty occurs on $T_k$ (with realization $t_k$). Let the Blue and Red casualties $\dot{X}_{i_t'}$ and $\dot{Y}_{i_t'}$ in a combat are r.v. whose densities are defined by $f_{S_X}(s/a_i, p_i, q_i)$ and $f_{S_Y}(s/b_i, p_i, q_i)$ respectively where forms of the densities are known except the unknown parameters $(a_i, b_i, p_i, q_i)$. It is assumed that the random sample $\left(\dot{X}_{i_t'}, \dot{Y}_{i_t'}\right)$ from $f_S(s)$ can be observed. On the basis of the observed sample values $\left(\dot{X}_{i_t'}, \dot{Y}_{i_t'}\right)$ it is desired to estimate the value of the unknown parameters $(a_i, b_i, p_i, q_i)$. *We further assume that the times between casualties are exponentially distributed reflecting a constant attrition rate consistent with Poisson process assumptions this is also supported by [16,32,33,34], then the pdf of casualty for the Red (X) and Blue (Y) sides associated to the equation (1) and (2) can be represented as in the equations (6) and (7). We have to note that while exponential is tractable and widely used, future work could explore weibull or mixed distribution for greater realism. Although exponential distribution enables Monte Carlo simulation without divergence, ensuring computational stability:*

$$f_{S_{X_{i'}}}(s) = \prod_{i=1}^{F}\left(a_i X_{i_t'}^{q_i} Y_{i_t}^{p_i}\right)\exp\left[-\left\{\sum_{i=1}^{F} a_i X_{i_t'}^{q_i} Y_{i_t}^{p_i}\right\} s\right] \tag{6}$$

$$-\infty < a_{i,}p_i, q_i < \infty, \ \forall i' = 1, 2, \ldots, F$$

$$f_{S_{Y_{i'}}}(s) = \prod_{i=1}^{F}\left(b_i X_{i_t}^{p_i} Y_{i_t'}^{q_i}\right) exp\left[-\left\{\sum_{i=1}^{F} b_i X_{i_t}^{p_i} Y_{i_t'}^{q_i}\right\} s\right]$$

$$-\infty < b_{i,}p_i, q_i < \infty, \forall i' = 1,2,\ldots,F \tag{7}$$

The likelihood equation of n pairs of random variable $\left(\dot{X}_{i_t'}, \dot{Y}_{i_t'}\right)$is defined as the joint density of the n pairs of random variables, which is considered to be a function of $(a_i, b_i, p_i, q_i)$. In particular, if $\left(\dot{X}_{i_t'}, \dot{Y}_{i_t'}\right)$ are independently and identically distributed random sample from the density $f_S(s)$, then the likelihood

function is:

$$f(X_{i_1'} \mid a_i, p_i, q_i) f(Y_{i_1'} \mid b_i, p_i, q_i), \ldots,$$
$$f(X_{i_t'} \mid a_i, p_i, q_i) f(Y_{i_t'} \mid b_i, p_i, q_i)$$

Then, the joint pdf will be:

$$L\left(a_i, b_i, p_i, q_i\right) = \prod_{t=1}^{N}\prod_{i=1}^{F}\left(a_i X_{i_t'}^{p_i} Y_{i_t}^{q_i}\right)^{\dot{X}_{i_t'}}\left(b_i Y_{i_t'}^{p_i} X_{i_t}^{q_i}\right)^{\dot{Y}_{i_t'}}$$
$$\exp\left\{-\left[\sum_{i=1}^{F} a_i X_{i_t'}^{p_i} Y_{i_t}^{q_i} + b_i Y_{i_t'}^{p_i} X_{i_t}^{q_i}\right] s\right\} \tag{8}$$

To construct the likelihood function from the available dataset, it is generally observed that casualty figures are generally available at daily interval. Let $L(a_i, b_i, p_i, q_i)$be the likelihood function for the random variables$\left(\dot{X}_{i_t'}, \dot{Y}_{i_t'}\right)$. If $\hat{a}_i, \hat{b}_i, \hat{p}_i, \hat{q}_i$are the values of $a_i, b_i, p_i, q_i$ which maximizes$L(a_i, b_i, p_i, q_i)$, then $\hat{a}_i, \hat{b}_i, \hat{p}_i, \hat{q}_i$ are the maximum-likelihood estimates of $a_i, b_i, p_i, q_i$.Now, instead of maximizing the likelihood function we will maximize its logarithmic form since both the maximum values occur at the same point and logarithmic form is easily computable.

Thus, on taking log of equation (8), we have

$$\ln L = \sum_{i=1}^{F}\sum_{t=1}^{N}\left[\begin{array}{l}\dot{X}_{i_t'} \ln\left(a_i X_{i_t'}^{p_i} Y_{i_t}^{q_i}\right) + \\ \dot{Y}_{i_t'} \ln\left(b_i Y_{i_t'}^{q_i} X_{i_t}^{p_i}\right) - \left(a_i X_{i_t'}^{p_i} Y_{i_t}^{q_i} + b_i Y_{i_t'}^{p_i} X_{i_t}^{q_i}\right) s\end{array}\right] \tag{9}$$

Differentiating the Log-likelihood function (9) partially with respect to $a_i$ and $b_i$ and equating it to zero, we have:

$$\frac{d \ln L}{d a_i} = \sum_{t=1}^{N}\frac{\dot{X}_{i_t'}}{a_i} - \sum_{t=1}^{N}\sum_{i=1}^{F} X_{i_t'}^{q_i} Y_{i_t}^{p_i} s = 0 \tag{10}$$

and

$$\frac{d\ln L}{db_i} = \sum_{t=1}^{N}\frac{\dot{Y}_{i'_t}}{b_i} - \sum_{t=1}^{N}\sum_{i=1}^{F} Y^{q_i}_{i'_t} X^{p_i}_{i_t} s = 0 \quad (11)$$

This gives

$$\sum_{t=1}^{N}\frac{\dot{X}_{i'_t}}{a_i} = \sum_{t=1}^{N}\sum_{i=1}^{F} X^{p_i}_{i'_t} Y^{q_i}_{i_t} s$$

and

$$\sum_{t=1}^{N}\frac{\dot{Y}_{i'_t}}{b_i} = \sum_{t=1}^{N}\sum_{i=1}^{F} Y^{p_i}_{i'_t} X^{q_i}_{i_t} s \qquad (12)$$

Thus, the maximum likelihood estimates are:

$$\hat{a}_i = \frac{\sum_{t=1}^{N}\dot{X}_{i'_t}}{\sum_{t=1}^{N}\sum_{i=1}^{F} X^{p_i}_{i'_t} Y^{q_i}_{i_t} s} \quad (13) \qquad \hat{b}_i = \frac{\sum_{t=1}^{N}\dot{Y}_{i'_t}}{\sum_{t=1}^{N}\sum_{i=1}^{F} Y^{p_i}_{i'_t} X^{q_i}_{i_t} s} \quad (14)$$

## 4. OVERVIEW OF KURSK DATA BASE

The Kursk Data Base (KDB) is developed by Dupuy Institute (DPI) [38] and is reformatted into a computerized database in1998. KDB is documented in the KOSAVE (Kursk Operation Simulation and Validation Exercise) [3]. The KDB contains daily on hand and losses for the four categories viz. manpower, tanks, APC and artillery for the Soviets and Germans for each of the 15 days of battle. Evidences of multiple force interaction in Kursk Battle shows multiple forces were fighting in the war. Therefore, developing heterogeneous model on this data is justified. In the present study, we have considered only the tank and artillery data for developing heterogeneous Lanchester model. Table 1 and 2 shows the tank and artillery weapons on hand and losses during the 14 days of battle. Figures 1–11 provide a progressive visualization of model performance. Figure 1 compares Soviet and German tank losses over the 14-day campaign. Figures 2 and 3 show fitted versus observed Soviet and German losses without phase division. Figures 4 and 5 present the corresponding fits using phased segments. Figures 6 and 7 decompose fitted losses into tank- and artillery-driven components under the heterogeneous model. Figures 8 and 9 display the log-likelihood contour and 3D surfaces (MLE, $p,q\in[-3,3]$). Figures 10 and 11 show the SSR 3D surface and contour plots (LSE, $p,q\in[-6,6]$) with corresponding a and b values.

This paper fits the generalized form of Heterogeneous Lanchester equations to the Battle of Kursk data using the method of Maximum Likelihood estimation and compares the performance of MLE with the techniques studied earlier such as the Sum of squared residuals (SSR), Linear regression and Newton-Raphson iteration. Different authors applied different methodologies for fitting Lanchester equations to the different battle data. The methodologies of Bracken, Fricker, and Clemen are applied to the Tank data of Battle of Kursk and results are shown in Table 3.

## 5. METHODOLOGY

First, we applied the technique of Least Square for estimating the parameters of the heterogeneous Lanchester model. The GRG algorithm is applied for maximizing the MLE and for minimizing the LSE. For implementing the Least Square approach, the Sum of Squared Residuals (SSR) is minimized. The expression of SSR for the equation (1) and (2) is given as:

$$SSR = \sum_{t=1}^{14}\left(\dot{X}_{i'_t} - \sum_{i=1}^{2} a_i X^{p_i}_{i'_t} Y^{q_i}_{i_t}\right)^2 + \sum_{t=1}^{14}\left(\dot{Y}_{i'_t} - \sum_{i=1}^{2} b_i Y^{p_i}_{i'_t} X^{q_i}_{i_t}\right)^2 \qquad (15)$$

For implementing this expression from Table 1 we have taken zero as initial values for all the unknown parameters. Then we start running the GRG algorithm iteratively. The GRG algorithm is available with the Microsoft Office Excel (2007) Solver [13]. The **GRG algorithm** is used because it efficiently solves the differentiable nonlinear optimization problems arising from the SSR and log-likelihood functions, updating parameters iteratively until convergence, with Excel Solver providing stable implementation using simple initial values and up to 1000 iterations for reliable estimates. The GRG solver uses iterative numerical method. The derivatives (and Gradients) play a crucial role in GRG. Once, we have the parameters we compute the estimated casualties. With the difference between the estimated and observed casualties, we computed the Sum of Squared Residuals. Similarly, we applied the GRG algorithm for optimizing the objective function as given in equation (9). We check the graphs of estimated and observed casualties for both the LS and MLE based approaches and found that if we divide the dataset into several subsets then we can improve the fit. Preliminary estimation showed that a single parameter set could not represent the full 15-day Kursk dynamics, producing systematic errors—consistent with earlier findings that **Lanchester coefficients** shift across battle stages as tactics and operational conditions change. To capture this temporal heterogeneity, the dataset (excluding the unrepresentative Day 1) is divided into five phases—($t_2$–$t_3$), ($t_4$–$t_6$), ($t_7$–$t_8$), ($t_9$–$t_{11}$), and ($t_{12}$–$t_{15}$)—with each treated as a short engagement. This phase-wise structure allows parameters to adjust to evolving conditions and consistently improves model fit. As we increase the number of divisions, the fit turns out to be better. The estimated casualty converges to the observed casualty. We have considered tank and artillery data for mixing the forces therefore $a_1$ (or $b_1$) represents effectiveness of Soviet (or Germans) tanks against Germans (or Soviets) tanks and $a_2$ (or $b_2$) represents effectiveness of Soviets (or Germans) Arty against Germans (or Soviets) tanks. The variation of attrition rates throughout battle tells us how the different player in the battle performs whether they are acting defensively or offensively.

The basic idea of using GRG algorithm is to quickly find optimal parameters that maximize the log-likelihood. The objective is to find the parameters that maximize the log-likelihood or in other words provide the best fit. Given the values in Table 1, we investigate what values of the parameters best fit the data. Although we derived the estimates for$a$ and $b$ using the MLE approach in equations (8) and (9), they are not applied directly. Log Likelihood is calculated using the equation (7) considering 0.5 as the initial value of the parameters. Then, we optimized the entire duration of the battle of the likelihood function using the GRG algorithm. The model obtained after estimation of parameters is:

$$\begin{aligned} \dot{X}_1 &= (1.46)X_1^{.129}Y_1^{.404} + (.906)X_1^{.138}Y_2^{.136} \\ \dot{Y}_1 &= (.704)Y_1^{.129}X_1^{.404} + (.953)Y_1^{.138}X_2^{.136} \end{aligned} \tag{16}$$

As the data for the first day is extremely low, we drop it since it will pose a problem in the computation of the likelihood and SSR function. Also, the extremely low casualty levels on the first day represent large outliers; thus, including the data of the first day affects the outcome to a great extent. Thus, the first day was dropped in fitting the data to the models. This approach is also justified by the historical account of the battle of Kursk, because the fight did not begin until July 5, the second day of the battle. Thus, dropping the data for the first day and dividing the remaining 14 days data into five phases, the total number of optimal parameters with each day as single phase is 102. Log-likelihood is calculated using equation (7) and is maximized separately for each of the five phases. Let $t$ denote the days, then the division is made as ($t_2$-$t_3$), ($t_4$-$t_6$), ($t_7$-$t_8$), ($t_9$-$t_{11}$), and ($t_{12}$-$t_{15}$). Fitting the model over multiple phases results in a better overall fit because there are additional parameters to explain the variation in casualties. The model has been improved from partitioning the battle into 14 phases. Each day of the battle is treated as mini-battle.

For the purpose of comparing models, $R^2$ value is calculated along with the Sum of squared residuals (SSR). $R^2$ value is calculated as:

$$R^2 = 1 - \frac{SSR}{SST} = 1 - \frac{\sum_{t=1}^{15}\left(\dot{X}_{i'_t} - \hat{\dot{X}}_{i'_t}\right)^2 + \sum_{t=1}^{15}\left(\dot{Y}_{i'_t} - \hat{\dot{Y}}_{i'_t}\right)^2}{\sum_{t=1}^{15}\left(\dot{X}_{i'_t} - \bar{\dot{X}}_{i'_t}\right)^2 + \sum_{t=1}^{15}\left(\dot{Y}_{i'_t} - \bar{\dot{Y}}_{i'_t}\right)^2} \tag{17}$$

A larger $R_2$ value indicates better fit. Also, Goodness-of-fit measures namely; Kolmogorov-Smirnov statistic [4] and Chi-square ( $\chi^2$ ) [4] have been calculated for the accuracy assessment of the MLE to that of the conventional approaches. Kolmogorov-Smirnov statistic is a measure of Goodness-of-fit, that is, the statistic tells us how well the model fits the observed data. The Kolmogorov-Smirnov (KS) statistic is based on the largest vertical difference between the theoretical and empirical (data) increasing distribution function.

$$KS = \max_{1 \le t^* \le 30} \left[ F\left(\hat{e}_{t^*}\right) - \frac{t^* - 1}{30}, \frac{t^*}{30} - F\left(\hat{e}_{t^*}\right) \right] \tag{18}$$

where $F\left(\hat{e}_{t^*}\right)$ is the cumulative distribution function of the estimated error between the observed losses and the estimated losses for both sides. Chi-Square ( $\chi^2$ ) is another measure of Goodness-of-fit. Chi-Square is given as:

$$\chi^2 = \sum_{t=1}^{15} \frac{\left(\dot{X}_{i_t'} - \hat{\dot{X}}_{i_t'}\right)^2}{\hat{\dot{X}}_{i_t'}} + \sum_{t=1}^{15} \frac{\left(\dot{Y}_{i_t'} - \hat{\dot{Y}}_{i_t'}\right)^2}{\hat{\dot{Y}}_{i_t'}} \tag{19}$$

where $\dot{X}_{i_t'}$ and $\hat{\dot{X}}_{i_t'}$ are the observed and expected casualties respectively.

For reproducibility, the heterogeneous Lanchester model estimation pipeline was implemented in Python, with the running code available as an open notebook on [42].

## 6. RESULTS AND DISCUSSIONS

Figures 2 and 3 shows the graphs of Soviet and German Tank losses along with the losses estimated through maximum likelihood approach. In this model a single set of parameters are estimated for representing the entire 14 days of the battle. Figures 4 and 5 show the performance of the same model when entire dataset is divided into 5 phases. From these figures, it is apparent that fitting the models with division into 5 phases resulted in a much better fit. Figures 6 and 7 show the further improvement in the dataset by dividing it into 14 phases where each day is considered as a mini battle. Further, the total losses are divided into two components: Losses due to tank and Losses due to Artillery. The overall SSR and likelihood values are functions of $p_i$'s and $q_i$'s. Figures 8 and 9 shows the 3D surfaces and contour plots of SSR as a function of $p_1$, $q_1$and $p_2$, $q_2$ respectively. From these figures, we can see that the minimum SSR zone is represented by contours of 1.5E+5 and 2.5E+5. Using a grid search in this zone, the best or optimal fit is obtained at $p_1$=.129, $q_1$=.404, $p_2$=.138, $q_2$=.136 with SSR 1.19E+5. The $a_1$, $b_1$, $a_2$, $b_2$ values corresponding to the optimal fit are 1.14, 0.70, 0.90, 0.95 respectively. Figures 10 and 11 shows the surface and contour plots of likelihood as a function of $p_1$, $q_1$ and $p_2$, $q_2$ respectively. From these figures, we can see that the zone of maximum likelihood is represented by contours of 6.0E+3 and 5.0E+3 with MLE 5.11E+3. Using a grid search in this zone, the best or optimal fit is obtained at $p_1$=.21, $q_1$=.28, $p_2$=.02, $q_2$=.04. The $a_1$, $b_1$, $a_2$, $b_2$ values corresponding to the optimal fit are 0.99, 0.88, 0.89, and 0.96 respectively.

Table 2 shows the results of Bracken [1], Fricker [7], Clemen [2] and MLE that have been applied on the tank versus tank and artillery data under heterogeneous situation. This table shows the KS statistic for MLE (with 14 divisions) is 0.08674, which is less than any other estimation methods implying that the method of MLE fits better as compared to the other methods. Also, $R_2$ is a measure of goodness of fit. The higher value of $R^2$ implies good fit to the data. The $R^2$ value of MLE (with 14 divisions) is 1. For comparing the efficiency of the different approaches, the root mean square error (RMSE) criterion is used. The RMSE of MLE with 5 divisions is 88.13 and the RMSE of MLE with 14 divisions is .0005, which is found to be the minimum. The RMSE of Clemen's Newton-Raphson Iteration model is 116.19, which is found to be the maximum. Therefore, efficiency ($E$) is measured with respect to the RMSE of the MLE with 14 divisions. Thus, the $E$ for MLE is maximum i.e. equal to 1 and E of Clemen's model is minimum i.e. equal to 4.30E-06. If the comparison is made among Bracken's, Fricker's and Clemens approaches, we can say that the Bracken approach is better. However, in all the cases the MLE outperforms other approaches. Based on all the GOF measures, it can be concluded that the MLE provides better fits.

In the present research we just demonstrate that if it is possible for mixing two forces it is also applicable for more than two forces. The number of parameters to be estimated increases fourth folded for mixing one additional force. With the estimated parameters, we computed the casualty due to tank component and casualty due to artillery component (See Table 3). When the 14 days Battle data is considered without any division, $a$ and $b$ parameters are significantly small and $a_1$>$b_1$which implies German tanks were more effective than Soviet tanks. Similarly, when we compare $a_2$ against $b_2$, $b_2$>$a_2$ that implies Soviet artillery were more effective than German artillery.

Table 3 shows the optimal parameters of heterogeneous Lanchester model with an $R^2$ of 1, RMSE of 0.0005, chi-square of 1.9E-5, SSR of 3.3E-6 and MLE of 13202. If we observe the residuals of the two sides it is seen that German casualty is more predictable than the Soviet. The smallest and largest residuals of Soviet side are obtained on the first (1.66E-08) and sixth (5.37E-07) day respectively. Similarly the smallest and largest residuals of German side are obtained on the fifth (7.21E-10) and fourteenth (9.07E-07) day respectively. Table 4 shows maximised log-likelihood values with divisions into 14 phases where each day is treated as a mini-battle. The parameters are obtained from maximum likelihood estimation of heterogeneous Lanchester model of tank and Artillery data (table 1) from Kursk Database with each day as single phase. Also, the parameter estimates $a_i$, $b_i$, $p_i$, $q_i$ are given corresponding to the maximised log-likelihood values with divisions. From this table we can see that the patterns of the parameters for each day of the battle are same for both the sides. In addition the tank component parameters are seen to be playing major role in the entire duration of the battle. Out of 14 days, 10 days the tank component parameters came out to be the maximum. That's why the result justified the Battle of Kursk and was correctly termed as the largest tank battle in the history. Table 5 shows the fitted tank losses and residual sum of square using heteregeneous Lanchester model.

Although dividing the battle into 14 mini-phases yields very high goodness-of-fit, penalized criteria and cross-validation confirm that the improvement is not solely due to overfitting. AIC and BIC values decrease substantially from the 1-phase to the 14-phase model, indicating that additional parameters provide explanatory power. However, the stronger penalty in BIC shows diminishing returns beyond 5 phases. Leave-one-day-out cross-validation further demonstrates that predictive accuracy remains stable across partitions, with the 14-phase model achieving the lowest error but only marginally better than the 5-phase model. This suggests that daily partitioning captures tactical variation but risks reduced generalizability, while 5-phase partitioning offers a balanced trade-off between fit and robustness.

To guard against overfitting, we applied leave-one-day-out cross-validation. For each day of the Kursk battle, the heterogeneous Lanchester model was fitted on the remaining 13 days and used to predict casualties for the omitted day. Prediction errors were aggregated using normalized mean squared error (NMSE). This procedure provides a robust measure of predictive accuracy.

Results (table 6) show that while the 14-phase model achieves the lowest NMSE, the improvement over the 5-phase model is marginal, suggesting that daily partitioning captures tactical variation but risks reduced generalizability. The 5-phase partition therefore offers a balanced trade-off between fit and robustness.

Beyond statistical fit, the estimated exponent parameters $p$ and $q$ provide insight into combat dynamics. Parameter $p$ reflects the scaling of offensive firepower with force size, while $q$ captures the sensitivity of defending forces to concentrated attack. In the Kursk data, tank engagements yielded moderate values of $p$ and $q$, indicating that concentration of armor produced disproportionate effects on losses. Artillery interactions, by contrast, showed very low exponents, consistent with area fire weapons where attrition scales linearly. These results suggest that heterogeneous engagements combine aimed fire dynamics with area fire dynamics, highlighting the importance of combined arms in shaping casualty outcomes.

## 7. CONCLUSIONS

Although mathematical formulations are well established for heterogeneous Lanchester model, very few studies have been done to model actual battle scenario. We have developed heterogeneous Lanchester model for Kursk Battle from World War II using tank and artillery data. All the previous studies on Kursk Battle were done to capture the homogeneous weapon system (Tank against Tank or Artillery against Artillery). The working principles of this model were only applicable for homogeneous situation. So extending those models in heterogeneous situation both theoretically and practically were main focus of this paper. We have formulated the likelihood expression under heterogeneous situation and applied to fit model under heterogeneous Lanchester model for Kursk database. We have estimated the MLE of the different parameters that are proved to be statistically more accurate. The unfamiliarity to deal with the heterogeneous situation by the previous approaches motivated us to venture the minute details of the Kursk Battle. The estimates are cross-validated to control the problem of the over fitting. Also, these estimates possess the optimal properties of consistency, sufficiency and efficiency. So compared to the previous work, the present paper opens up the opportunity for exploring the complicated structure of Kursk Battle of World War II.

**Table 1* Day-by-day Soviet and German tank and artillery holdings and casualties**

| Days | Soviet's Tank On Hand | Soviet's Tank Losses | German's Tank On Hand | German's Tank Losses | Soviet's Arty On Hand | Soviet's Arty Losses | German's Arty On Hand | German's Arty Losses |
|---|---|---|---|---|---|---|---|---|
| 1 | 2396 | 105 | 986 | 198 | 705 | 13 | 1166 | 24 |
| 2 | 2367 | 117 | 749 | 248 | 676 | 30 | 1161 | 5 |
| 3 | 2064 | 259 | 673 | 121 | 661 | 15 | 1154 | 7 |
| 4 | 1754 | 315 | 596 | 108 | 648 | 14 | 1213 | 13 |
| 5 | 1495 | 289 | 490 | 139 | 640 | 9 | 1210 | 6 |
| 6 | 1406 | 157 | 548 | 36 | 629 | 13 | 1199 | 12 |
| 7 | 1351 | 135 | 563 | 63 | 628 | 7 | 1206 | 15 |
| 8 | 977 | 414 | 500 | 98 | 613 | 16 | 1194 | 12 |
| 9 | 978 | 117 | 495 | 57 | 606 | 10 | 1187 | 7 |
| 10 | 907 | 118 | 480 | 46 | 603 | 5 | 1184 | 5 |
| 11 | 883 | 96 | 426 | 79 | 601 | 5 | 1183 | 3 |
| 12 | 985 | 27 | 495 | 23 | 600 | 3 | 1179 | 4 |
| 13 | 978 | 42 | 557 | 7 | 602 | 0 | 1182 | 2 |
| 14 | 948 | 85 | 588 | 6 | 591 | 4 | 1182 | 11 |

*Database of the Battle of Kursk of WW II [3]

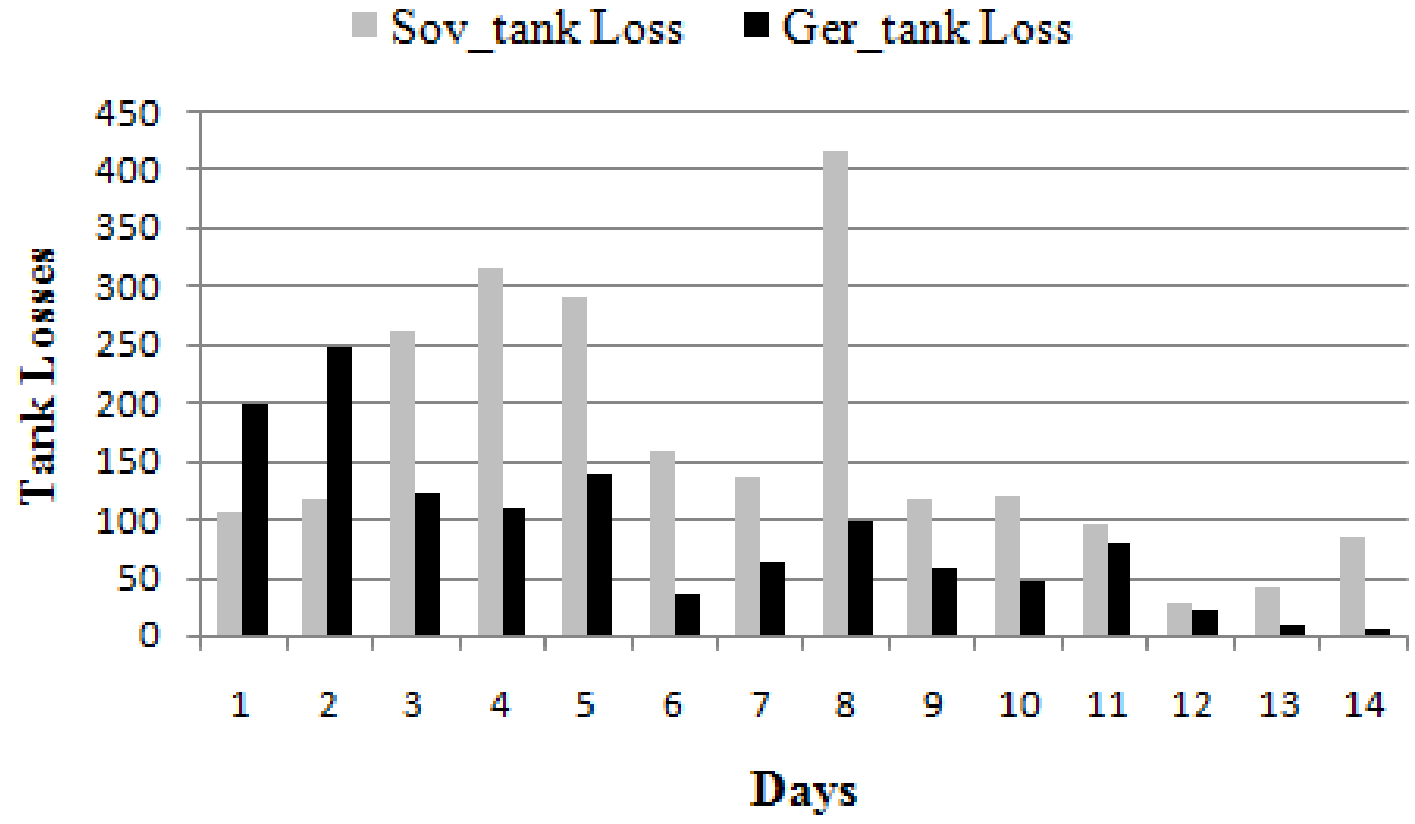

**Figure 1. Comparison of daily tank losses for Soviet and German forces during the Kursk campaign.**

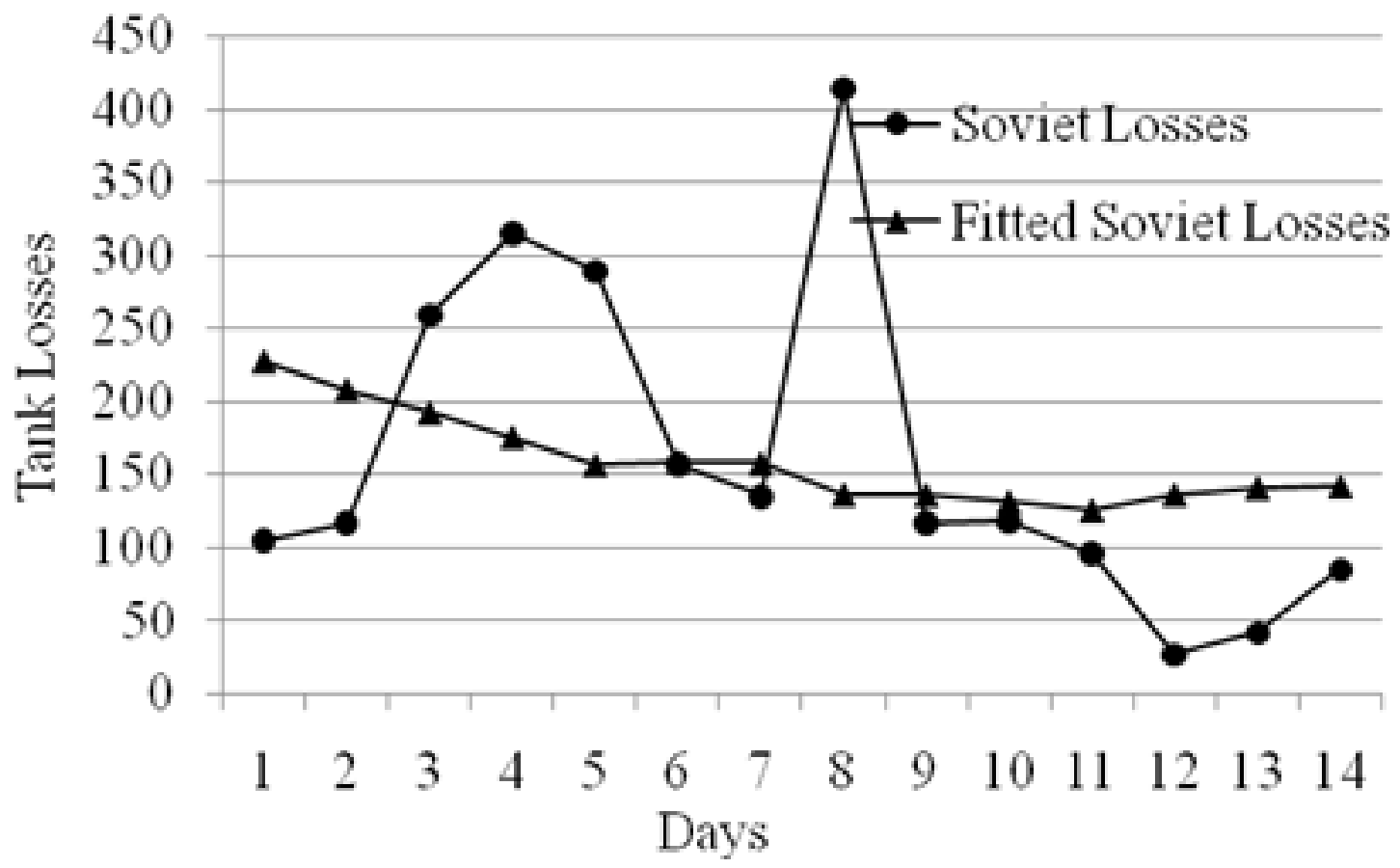

**Figure 2**: **Fitted versus observed daily Soviet tank losses for the Kursk campaign without phase division.**

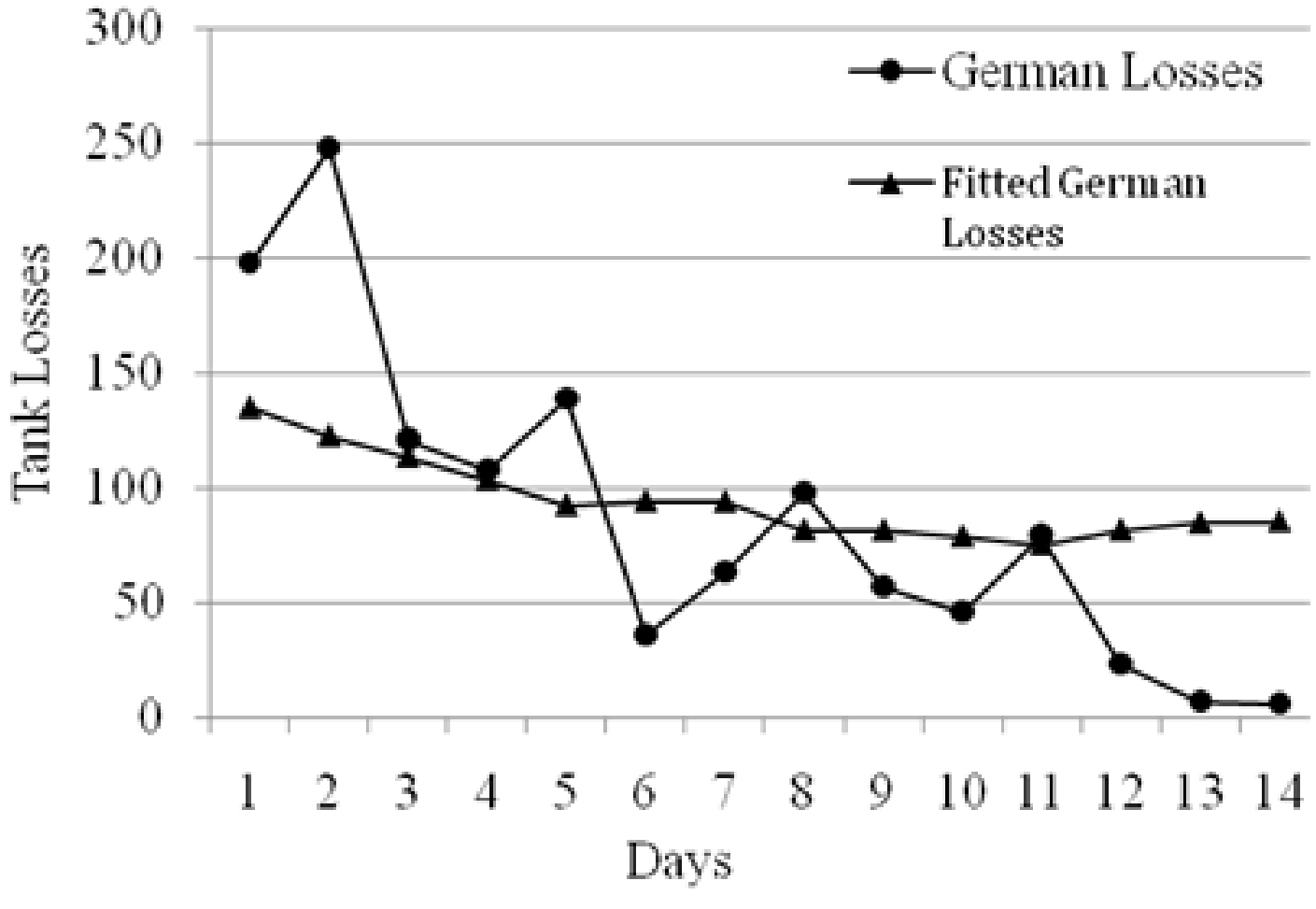

**Figure 3**: **Fitted versus observed daily German tank losses for the Kursk campaign without phase division.**

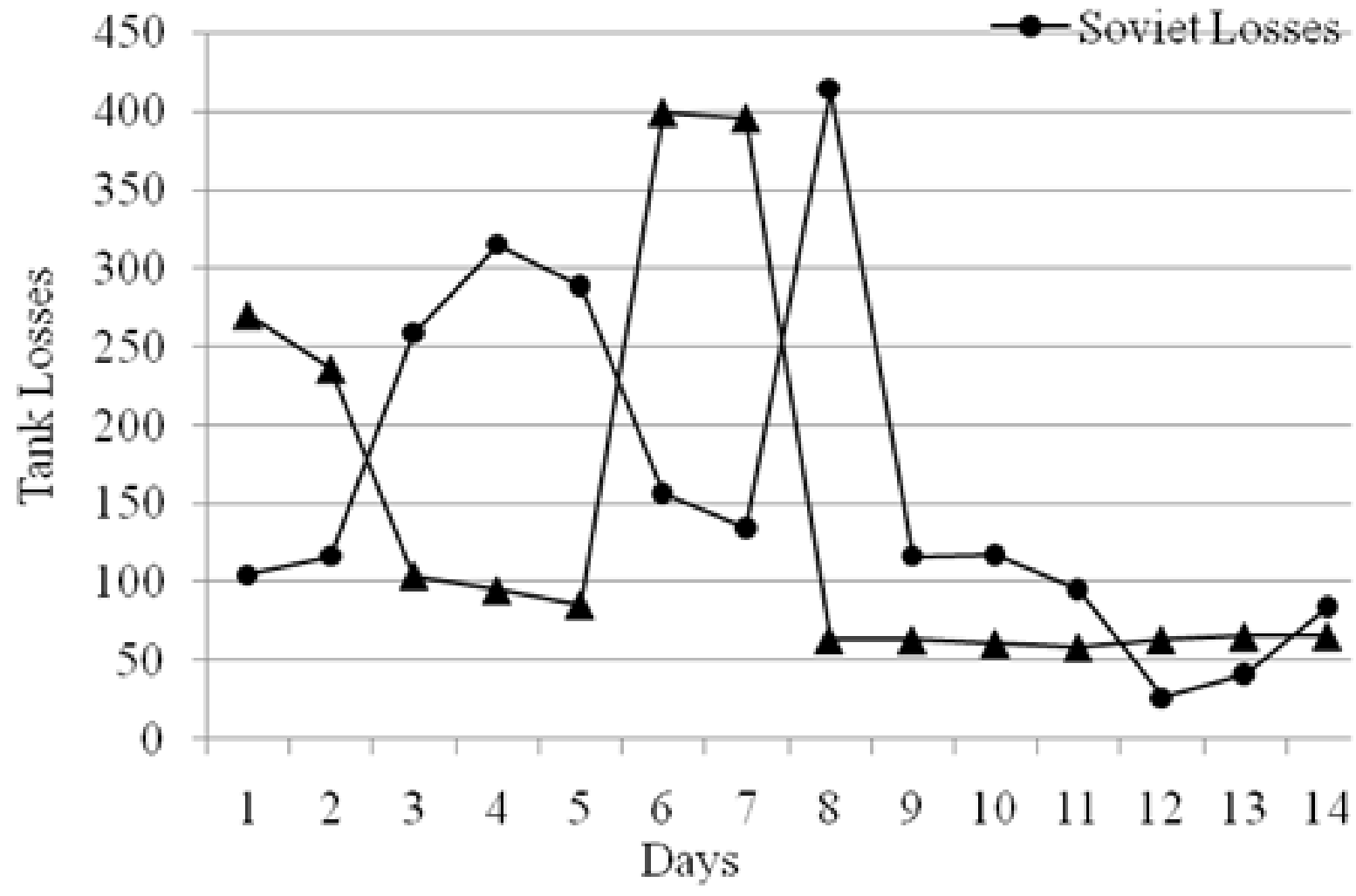

**Figure 4**: **Fitted versus observed Soviet tank losses across phased segments of the Kursk campaign.**

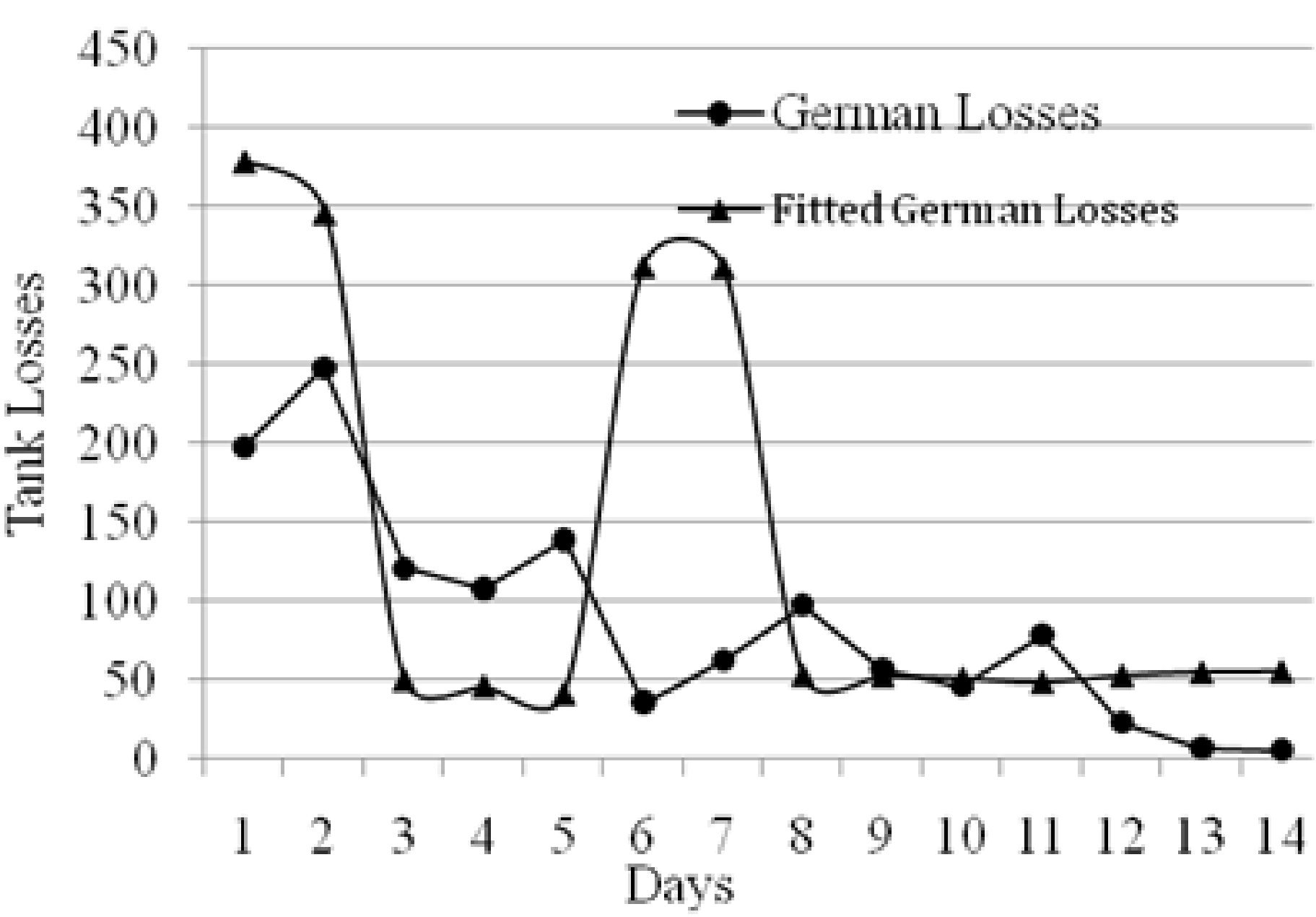

**Figure 5**: **Fitted versus observed German tank losses across phased segments of the Kursk campaign.**

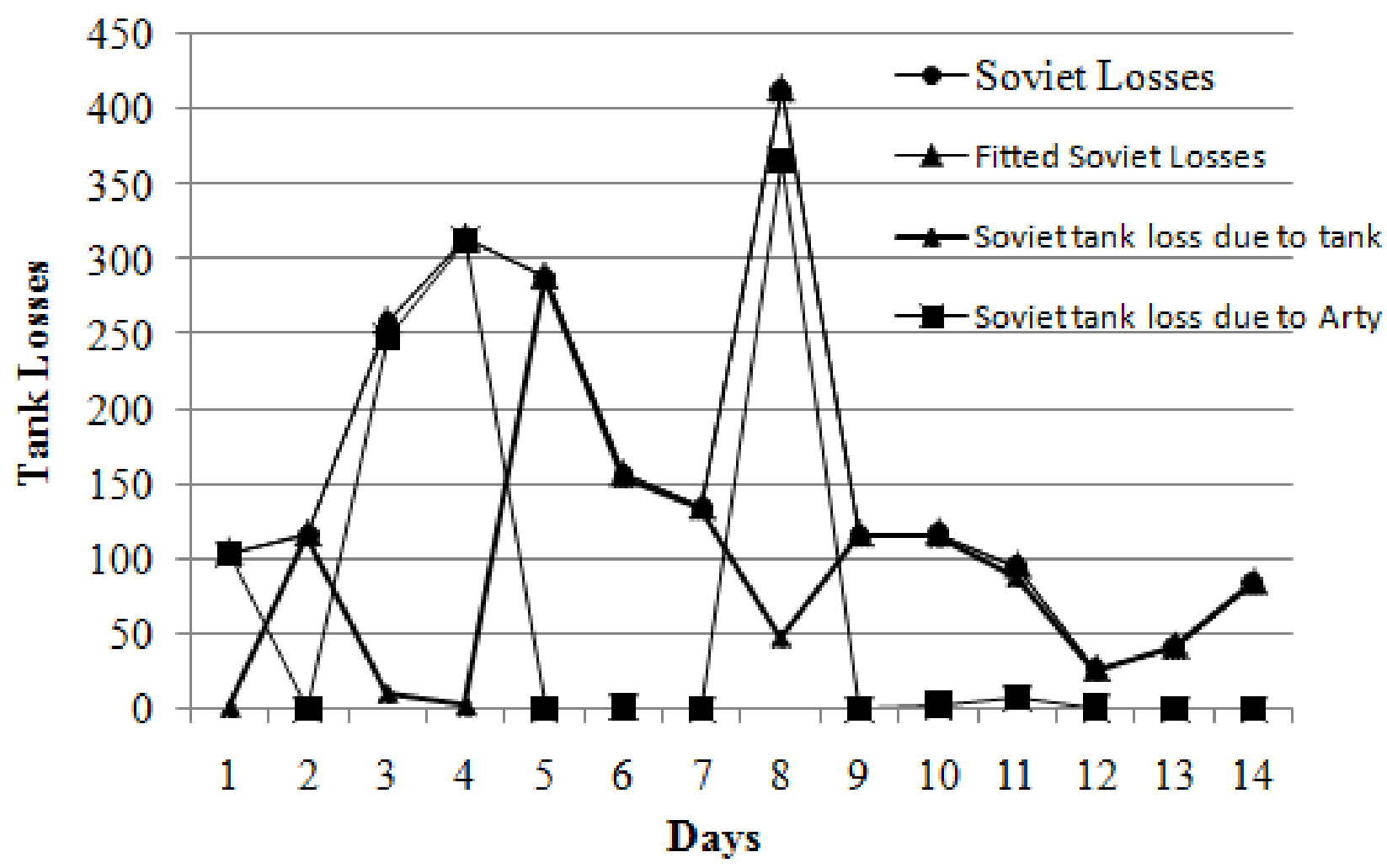

**Figure 6: Fitted Soviet tank losses from German tank and artillery fire under the heterogeneous Lanchester model, with**

**total losses decomposed into tank- and artillery-driven components.**

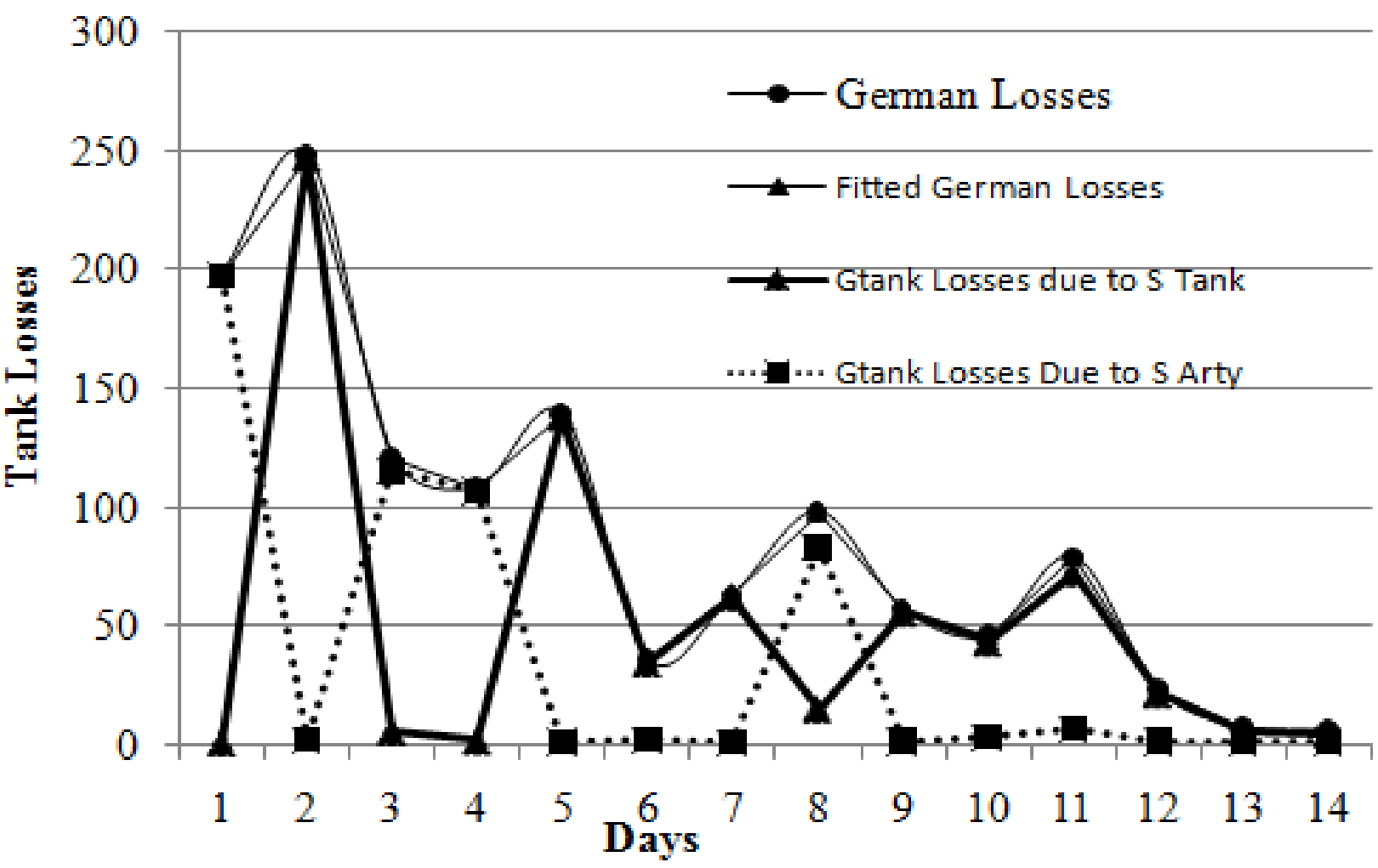

**Figure 7: Fitted versus observed German tank losses under the heterogeneous Lanchester model, with total losses decomposed into tank- and artillery-driven components.**

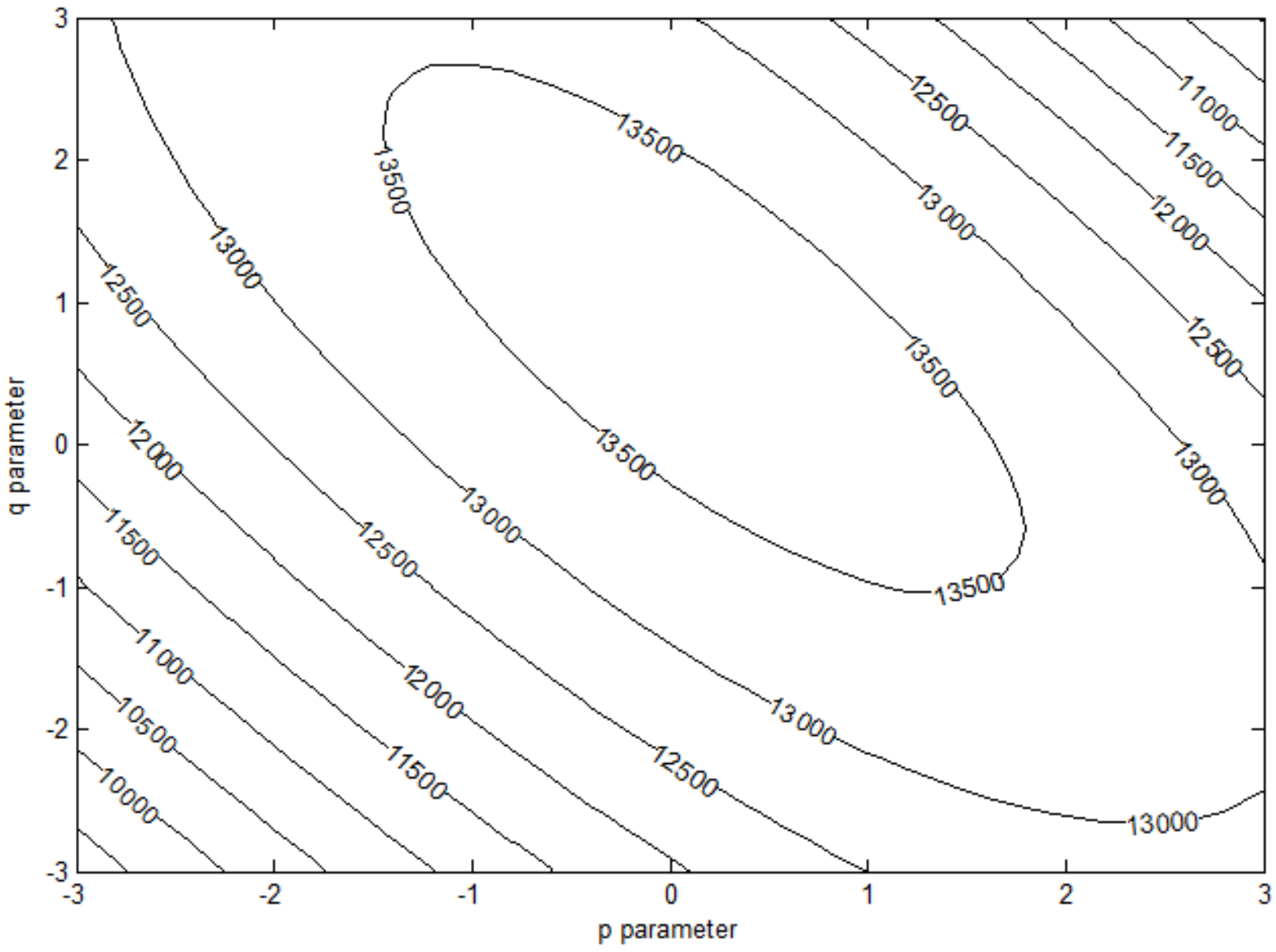

**Figure 8: Contour plot of log-likelihood values for Soviet and German tank data across the Kursk campaign, with p and q varied from −3 to 3 under MLE-based parameter estimation.**

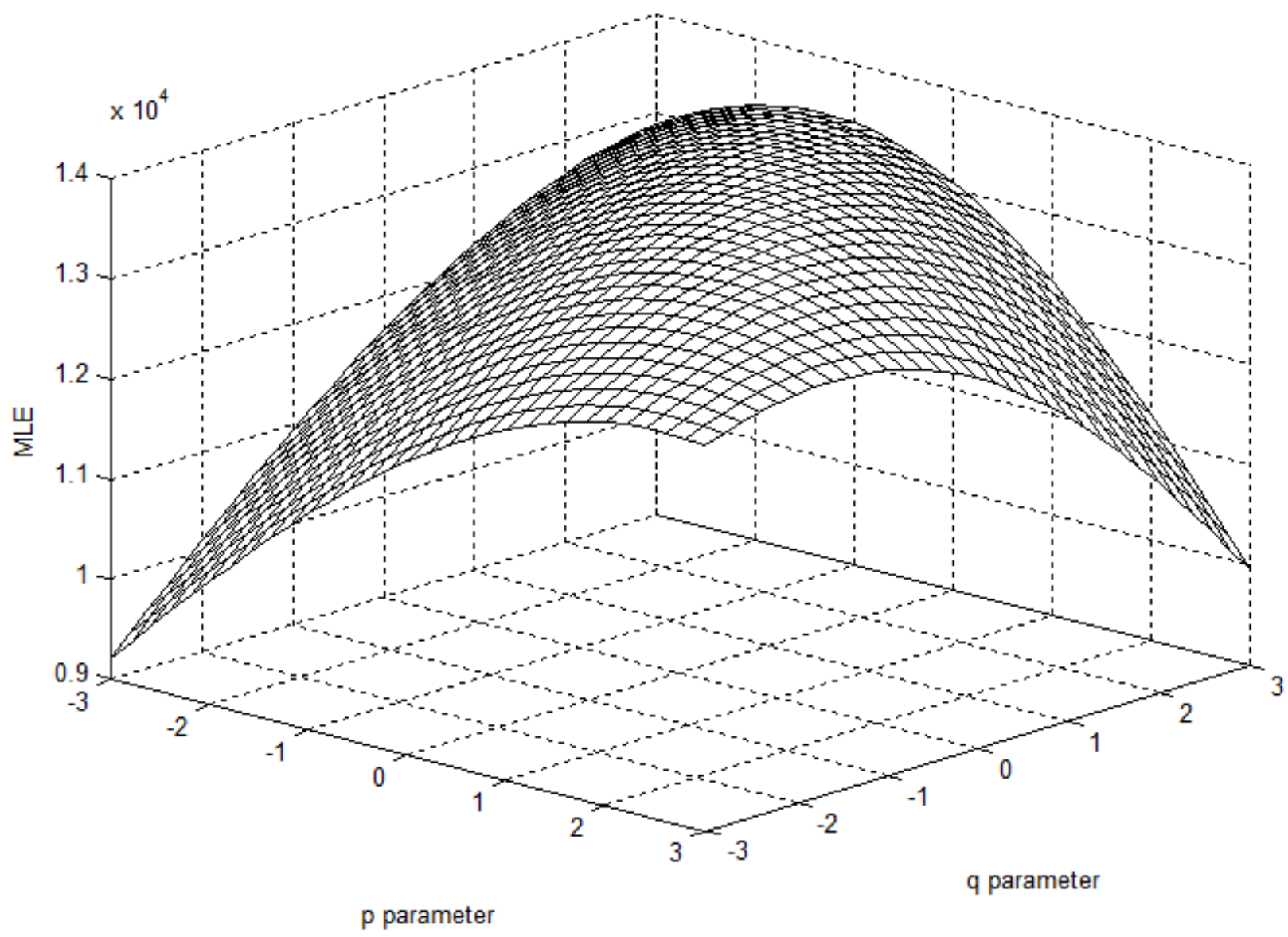

**Figure 9: 3D log-likelihood surface for Kursk tank data, with p and q varied from −3 to 3 and corresponding a and b values estimated via the MLE approach.**

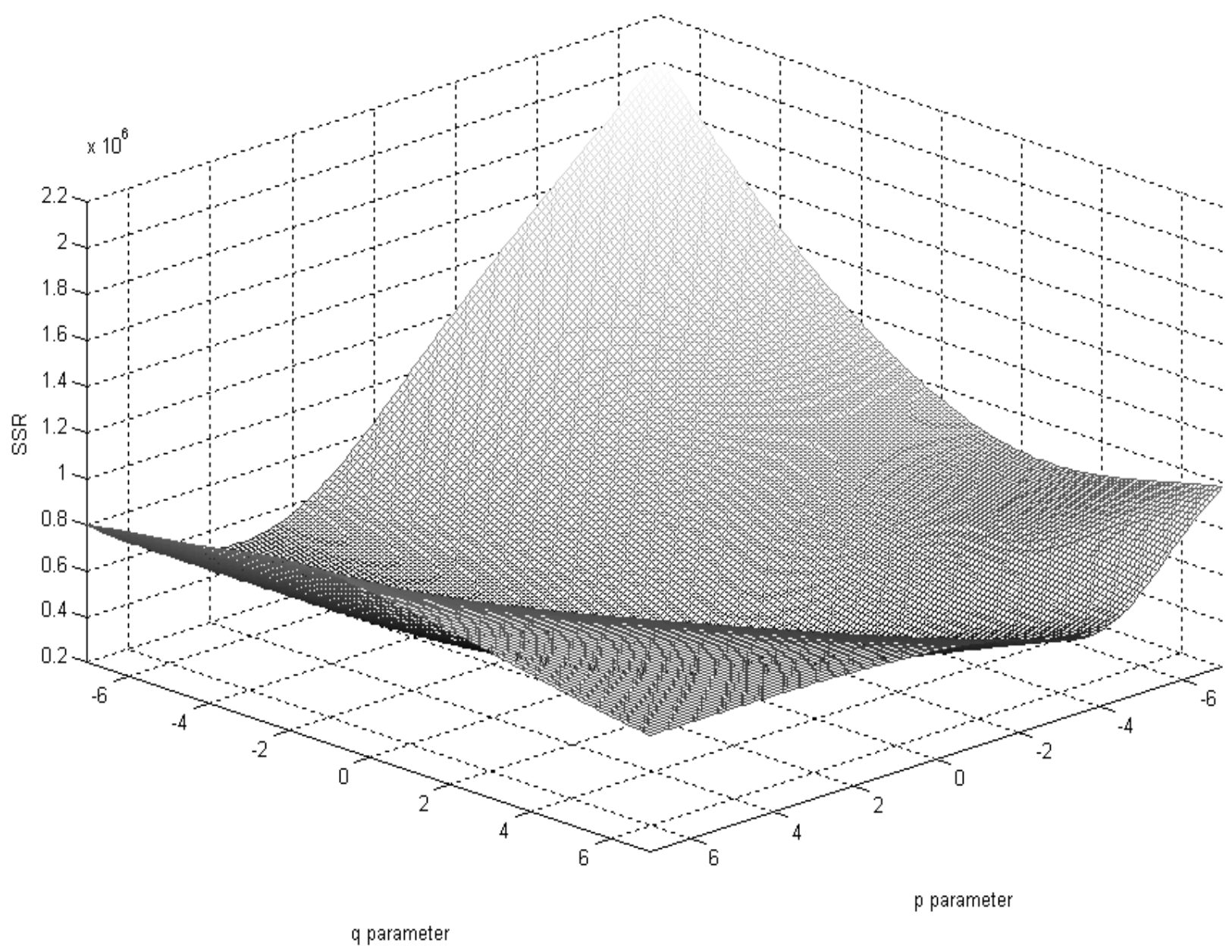

**Figure 10: 3D SSR surface for Kursk tank data, with p and q varied from −6 to 6 and corresponding a and b values obtained via the Least Squares approach.**

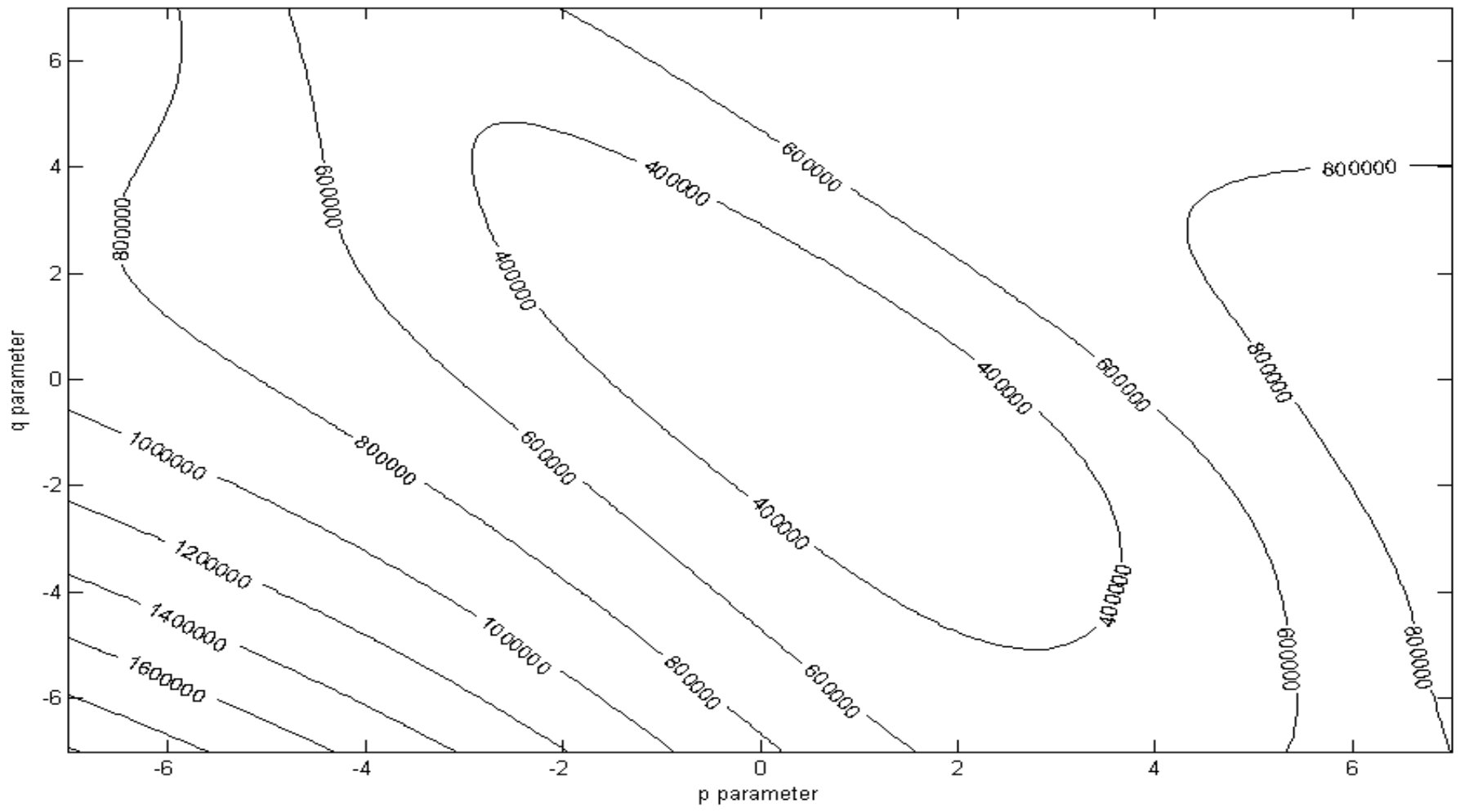

**Figure 11**: **Contour plot of SSR for Soviet and German tank data across the Kursk campaign, with p and q varied from**

**−6 to 6 and parameters estimated via the Least Squares approach.**

Table 2. Comparison of different estimation methods with respect to common Goodness-of-fit measures such as sum of square residuals (*SSR*), Kolmogorov-Smirnov (*KS*) Statistic, Chi-Square ($\chi^2$) and *R*-square ($R^2$), root mean square error (*RMSE*) and efficiency(*E*) from homogeneous tank against tank data of the Kursk battle.

| Sl. No. | Approaches | SSR/log-Likelihood | *KS* | $\chi^2$ | $R^2$ | *RMSE* | *E** |
|---|---|---|---|---|---|---|---|
| 1 | Bracken [1] | | | | | | |
| | Model 1 | 1.19E+5 | 0.1567 | 2954 | 0.48 | 92.19 | 0.9559 |
| 2 | Fricker [6] | | | | | | |
| | Model 1 | 1.29E+5 | 0.1065 | 3082 | 0.44 | 95.99 | 0.9181 |
| 3 | Clemens [2] | | | | | | |
| | Linear Regression | 1.88E+5 | 0.1063 | 3854 | 0.22 | 115.88 | 0.7605 |
| | Newton-Raphson Iteration | 1.89E+5 | 0.1123 | 3520 | 0.22 | 116.19 | 0.7584 |
| | MLE(Log-Likelihood) | | | | | | |
| 4 | Without division | 13203 | 0.1053 | 2580 | 0.71 | 89.72 | 0.9822 |
| 5 | With Division (4 phases) | 13313 | 0.0909 | 2670 | 0.82 | 88.13 | 1 |

*Efficiency is calculated with respect to the RMSE of MLE (with division)

Table 3. Maximization of likelihood estimation with divisions from homogeneous tank against tank data of Kursk battle.

| | Likelihood | *a* | *b* | *p* | *q* |
|---|---|---|---|---|---|
| Phase 1 | 2621.35 | 0.9031 | 1.0887 | 0.4793 | 0.3079 |
| Phase 2 | 6169.72 | 1.1327 | 0.7884 | 0.1418 | 0.4720 |
| Phase 3 | 2354.21 | 1.0300 | 0.8816 | 0.3889 | 0.4838 |
| Phase 4 | 2168.54 | 1.0221 | 0.8939 | 0.2793 | 0.3483 |

Table 4 The Parameters of Heterogeneous Lanchester Model. The parameters are obtained from maximum likelihood estimation from heterogeneous tank against tank and artillery data of Kursk battle with each day as single phase.

| | Likelihood | $a_1$ | $a_2$ | $b_1$ | $b_2$ | $p_1$ | $p_2$ | $q_1$ | $q_2$ |
|---|---|---|---|---|---|---|---|---|---|
| Phase I | 1232.74 | 0.929 | 0.456 | 0.935 | 1.395 | 0.011 | 0.467 | 5E-11 | 0.273 |
| Phase II | 1559.50 | 0.795 | 0.670 | 1.119 | 1.188 | 0.539 | 0.039 | 0.182 | 0.000 |
| Phase III | 1639.50 | 1.062 | 1.021 | 0.808 | 0.887 | 0.015 | 0.371 | 0.285 | 0.377 |
| Phase IV | 1894.73 | 0.989 | 1.133 | 0.870 | 0.753 | 0.000 | 0.352 | 0.113 | 0.418 |
| Phase V | 1895.48 | 1.140 | 0.882 | 0.815 | 0.979 | 0.215 | 0.024 | 0.574 | 0.000 |
| Phase VI | 729.83 | 1.301 | 0.897 | 0.534 | 0.967 | 0.000 | 0.055 | 0.660 | 0.043 |
| Phase VII | 725.22 | 1.151 | 0.886 | 0.724 | 0.972 | 0.164 | 0.011 | 0.516 | 0.000 |
| Phase VIII | 2432.03 | 1.290 | 1.332 | 0.568 | 0.516 | 0.020 | 0.388 | 0.507 | 0.416 |
| Phase IX | 613.62 | 1.196 | 0.889 | 0.714 | 0.967 | 0.184 | 0.000 | 0.498 | 0.030 |
| Phase X | 575.05 | 1.270 | 0.906 | 0.612 | 0.960 | 0.140 | 0.072 | 0.535 | 0.106 |
| Phase XI | 608.36 | 1.017 | 0.904 | 0.888 | 0.972 | 0.300 | 0.138 | 0.390 | 0.176 |
| Phase XII | 111.10 | 0.996 | 0.891 | 0.885 | 0.968 | 0.210 | 0.022 | 0.282 | 0.048 |
| Phase XIII | 121.60 | 1.369 | 0.907 | 0.266 | 0.936 | 0.002 | 0.000 | 0.492 | 0.000 |
| Phase XIV | 297.37 | 1.550 | 0.911 | 0.119 | 0.945 | 0.007 | 0.005 | 0.576 | 0.018 |

**Table 5. Fitted Tank Losses and residual sum of square using Heteregeneous Lanchester model. The tank and arty components of the fitted models are obtained through maximum likelihood estimation method from heterogeneous tank against tank and artillery data of Kursk battle.**

| Days | Soviet | | | | German | | | |
|---|---|---|---|---|---|---|---|---|
| | SLoss-Fit | STank Comp. | SArty Comp. | SResid-ual^2 | GLoss-Fit | GTank Comp. | GArty Comp. | GResid-ual^2 |
| 1 | 105.00 | 1.01 | 103.99 | 1.66E-08 | 198.00 | 1.02 | 196.98 | 3.72E-08 |
| 2 | 117.00 | 116.12 | 0.88 | 1.82E-08 | 248.00 | 246.47 | 1.53 | 1.34E-07 |
| 3 | 259.00 | 10.34 | 248.66 | 9.09E-08 | 121.00 | 5.81 | 115.19 | 3.33E-08 |
| 4 | 315.00 | 2.29 | 312.71 | 1.1E-07 | 108.00 | 1.79 | 106.21 | 3.9E-08 |
| 5 | 289.00 | 287.95 | 1.04 | 1.03E-07 | 139.00 | 137.86 | 1.14 | 7.21E-10 |
| 6 | 157.00 | 155.19 | 1.81 | 5.37E-07 | 36.00 | 34.20 | 1.81 | 3.48E-07 |
| 7 | 135.00 | 134.04 | 0.96 | 8.65E-08 | 63.00 | 61.96 | 1.04 | 2.05E-08 |
| 8 | 414.00 | 47.77 | 366.23 | 1.4E-07 | 98.00 | 15.18 | 82.82 | 4.39E-08 |
| 9 | 117.00 | 115.91 | 1.09 | 2.76E-08 | 57.00 | 55.83 | 1.17 | 1.32E-08 |
| 10 | 118.00 | 114.91 | 3.09 | 2.68E-08 | 46.00 | 43.08 | 2.92 | 1.06E-08 |
| 11 | 96.00 | 88.11 | 7.89 | 4.33E-08 | 79.00 | 72.21 | 6.79 | 7.96E-10 |
| 12 | 27.00 | 25.55 | 1.45 | 2.02E-08 | 23.00 | 21.50 | 1.50 | 9.59E-09 |
| 13 | 42.00 | 41.09 | 0.91 | 2.43E-07 | 7.00 | 6.06 | 0.94 | 5.48E-09 |
| 14 | 85.00 | 83.94 | 1.06 | 2.64E-07 | 6.00 | 4.91 | 1.09 | 9.07E-07 |

**Table 6. Model Comparison**

| Model Partition | Parameters (k) | Log-Likelihood | AIC | BIC | CV Error (NMSE) |
|---|---|---|---|---|---|
| 1-phase | 4 | 2621.35 | -5234.7 | -5232.1 | 0.42 |
| 5-phase | 20 | 6169.72 | -12299.4 | -12286.0 | 0.28 |
| 14-phase | 56 | 2354.21 | -4596.4 | -4558.8 | 0.25 |
| Mixed-phase | 10 | 2168.54 | -4317.1 | -4310.7 | 0.30 |

$AIC=2k-2ln(n)$, $BIC=kln(n)-2ln(L)$, NMSE=average of sum of squared casualty difference w.r.t variance

**Abbreviations**

**LSE** — Least Square Estimation
**MLE** — Maximum Likelihood Estimation
**SSR** — Sum of Squared Residuals
**GOF** — Goodness-of-Fit
**KS** — Kolmogorov–Smirnov
**RMSE** — Root Mean Square Error
**KDB** — Kursk Data Base
**DPI** — Dupuy Institute
**KOSAVE** — Kursk Operation Simulation and Validation Exercise
**APC** — Armored Personnel Carrier
**Arty** — Artillery

**ISSA** — Institute For Systems Studies And Analyses
**DRDO** —Defence Research And Development Organization
**WW II** — World War II

**Author Contributions**

Sumanta Kumar Das: Model formulation; data preparation; parameter estimation using LSE and MLE; implementation of GRG optimization; generation of figures and statistical analyses; interpretation of results; manuscript writing and revision.

**Data Availability Statement**

The data that support the findings of this study are openly available in TDI (The Dupuy Institute) at https://www.dupuyinstitute.org/data/kursk.htm, reference number [38].

**Funding**

No funding was received

**Conflicts of Interest**

The author declares no conflicts of interest related to the research, authorship, or publication of this work.